\documentclass[aps,prd,twocolumn,superscriptaddress]{revtex4}
\usepackage{epsfig,epsf}
\usepackage{amsmath}
\usepackage{amsthm}
\usepackage{amsfonts}
\usepackage{amssymb}
\usepackage{dsfont}
\usepackage{multirow}
\usepackage{appendix}
\usepackage{slashed}
\usepackage[active]{srcltx}
\usepackage{psfrag}

\begin{document}

\title{Double strange hybrid baryon }
\date{\today}
\author{B.~Barsbay}
\affiliation{Division of Optometry, School of Medical Services and Techniques, Dogus University, Dudullu-\"{U}mraniye, 34775
Istanbul, T\"{u}rkiye}
\author{K.~Azizi}
\thanks{Corresponding author: kazem.azizi@ut.ac.ir}
\affiliation{Department of Physics, University of Tehran, North Karegar Avenue, Tehran
14395-547, Iran}
\affiliation{Department of Physics, Faculty of Engineering and Natural Sciences, Dogus University, Dudullu-\"{U}mraniye, 34775
Istanbul, T\"{u}rkiye}
\author{H.~Sundu}
\affiliation{Department of Physics Engineering, Istanbul Medeniyet University, 34700
Istanbul, T\"{u}rkiye}

\begin{abstract}
We investigate the spectroscopic properties of the double strange hybrid baryon with quark 
content $ssqg$ within the framework of QCD sum rule. Using an interpolating current with 
explicit gluonic degrees of freedom, the two-point correlation function is analyzed in terms 
of two independent Lorentz structures, $\slashed{q}$ and $I$. The operator product expansion 
is carried out by including vacuum condensates up to dimension ten, and the corresponding 
sum rules are derived for both structures. By taking the average of the results obtained 
from the two Lorentz structures, we extract the masses and pole residues of the ground and 
first excited states. For the ground state, we obtain a mass of
$\widetilde{
M } = (1593.44\pm 130.29)~ \mathrm{MeV}$ and a residue of $\widetilde{\lambda } = (2.57\pm 0.40) 
\times10^{-3} ~\mathrm{GeV}^5$. For the first excited state, the corresponding values are 
$M = (1897.47\pm 124.44)~ \mathrm{MeV}$ and 
$\lambda = (2.88\pm 0.62) \times10^{-3} ~\mathrm{GeV}^5$. The obtained results provide 
theoretical predictions for the double strange hybrid baryon spectrum and may be useful for 
future experimental searches as well as further nonperturbative studies of hybrid hadrons.
\end{abstract}

\maketitle


\section{Introduction}

\label{sec:Intro}

The investigation of hadronic states beyond the conventional quark model has attracted 
considerable attention in hadron spectroscopy. Within the framework of quantum 
chromodynamics (QCD), hadrons are not restricted to conventional mesons ($q\bar{q}$) and 
baryons ($qqq$); rather, QCD predicts a rich spectrum of exotic configurations, including 
tetraquarks, pentaquarks, glueballs, and hybrid hadrons with explicit gluonic degrees of 
freedom \cite{Olsen:2014mea}. Among these exotic states, hybrid baryons are of particular 
interest, as they provide a unique laboratory for studying the dynamical role of gluonic 
excitations in baryonic systems. The experimental observation and characterization of hybrid 
baryons would provide important insights into the nonperturbative dynamics of QCD and the 
mechanism of color confinement. Therefore, reliable theoretical predictions for the 
spectroscopic properties of hybrid baryons are essential for guiding future experimental 
searches and for improving our understanding of the gluonic structure of hadrons.

In recent decades, hybrid mesons have been extensively investigated within a variety of theoretical approaches~\cite{Isgur:1985vy,Isgur:1984bm,Burns:2006wz,Chanowitz:1982qj,Barnes:1982tx,Lacock:1996vy,Lacock:1996ny,Lacock:1997an,Bernard:2003jd,Hedditch:2005zf,Dudek:2009qf,Dudek:2010wm,Dudek:2011tt,Dudek:2013yja,Ma:2020bex,Balitsky:1982ps,Govaerts:1983ka,Govaerts:1984bk,Govaerts:1984hc,Balitsky:1986hf,Govaerts:1986pp,Jin:2002rw,Berg:2012gd,Chen:2013zia,Chen:2013eha,Palameta:2018yce,Ho:2018cat,Ho:2019org,Barsbay:2022gtu,Li:2021fwk,Chen:2022qpd,Wang:2023whb,Wang:2025ypo,Barsbay:2024vjt,Alaakol:2024zyh,Esmer:2025xss,Agaev:2025llz,Barsbay:2025vjq}. These studies have led to several experimentally observed candidates for light hybrid mesons, such as $\pi_1(1400)$~\cite{IHEP-Brussels-LosAlamos-AnnecyLAPP:1988iqi}, $\pi_1(1600)$~\cite{E852:2001ikk}, $\pi_1(2015)$~\cite{E852:2004gpn}, and $\eta_1(1855)$~\cite{BESIII:2022riz,BESIII:2022iwi}, which are characterized by exotic quantum numbers $J^{PC}=1^{-+}$. By contrast, hybrid baryons are generally described as consisting of a color-octet three-quark configuration coupled to a color-octet gluonic excitation. However, their theoretical and experimental investigations remain far less developed than those of hybrid mesons. A major difficulty arises from the fact that hybrid baryons do not possess manifestly exotic quantum numbers and can therefore mix strongly with conventional excited $qqq$ baryon states carrying the same quantum numbers. Such mixing between $qqqg$ hybrid configurations and ordinary baryon excitations is naturally allowed in QCD, complicating the unambiguous identification of hybrid baryon states.

Consequently, unlike hybrid mesons with exotic $J^{PC}$ assignments, hybrid baryons are expected to appear as additional resonances in the ordinary baryon spectrum rather than as states with unique quantum numbers. As a result, their strong mixing with conventional baryons makes the extraction of genuine hybrid signals particularly difficult, highlighting the importance of reliable theoretical predictions for interpreting experimental data and guiding future experimental searches.

To shed light on their nature and properties, hybrid baryons have been studied using a variety of theoretical approaches, including the constituent quark model~\cite{Burkert:2017djo,Viseur:2026tma}, relativistic quark model~\cite{Gerasyuta:2002hg}, potential model~\cite{Cimino:2024bol}, flux-tube model~\cite{Capstick:1999qq,Capstick:2002wm}, MIT bag model~\cite{Barnes:1982fj,Golowich:1982kx}, lattice QCD~\cite{Dudek:2012ag,Khan:2020ahz}, and QCD sum rules ~\cite{Martynenko:1991pc,Kisslinger:1995yw,Kisslinger:2003hk,Azizi:2017xyx,Zhao:2023imq,Wang:2024lnv,Yang:2025hzc}. Although these studies provide valuable information on the mass spectrum and internal structure of hybrid baryons, many of their properties remain uncertain. Therefore, further investigations employing complementary nonperturbative techniques are necessary to achieve a more comprehensive understanding of these states.

Motivated by the need for further theoretical studies of hybrid baryons, in this work we 
investigate the spectroscopic properties of the double strange hybrid baryon with quark 
content $ssqg$ within the framework of QCD sum rules. We determine the masses and pole 
residues of both the ground and first excited states, providing theoretical predictions that 
may be useful for future experimental searches and for a better understanding of strange 
hybrid baryon spectroscopy.

The remaining of this work is organized as follows. In Sec.~\ref{sec:ssqgmass}, we construct 
the interpolating current for the double strange hybrid baryon and derive the corresponding 
QCD sum rules for the masses and residues. Section~\ref{sec:Numeric} is devoted to the 
numerical analysis, where the spectroscopic parameters of the ground and first excited 
states are extracted. Finally, Sec.~\ref{sec:Dis} contains our conclusions and discussion.


\section{ Spectroscopic Parameters of Double Strange Hybrid Baryons }

\label{sec:ssqgmass}
In this section, we investigate the spectroscopic properties of the double strange hybrid baryon with quark-gluon content $ssqg$ within the framework of QCD sum rules. We consider both the negative-parity ground state and the orbitally excited positive-parity state, which couple to the same interpolating current. To determine the masses and pole residues of these states, we begin with the following two-point correlation function:
\begin{equation}
\Pi_{H}(q)=i\int d^{4}xe^{iq\cdot x}\langle 0|\mathcal{T}\lbrace \eta_{H}(x)\bar{\eta}_{H}(0)\rbrace|0\rangle ,  \label{eq:CorrF1}
\end{equation}
where ${\eta}_{H}$ is the interpolating current for the double strange hybrid baryon. The explicit form of the interpolating current is chosen as
\begin{eqnarray}
\eta_{H}(x)&=& g_{s} \epsilon^{abc}\left[ s^a(x)C\gamma^{\mu} s^b(x)\right]  \gamma^{\nu} \gamma_{5} \left[G_{\mu \nu} q(x)\right]^c ,\label{eq:current} 
\end{eqnarray}
where $a$, $b$, and $c$ denote color indices, $C$ is the charge-conjugation matrix, and the quark fields are represented by $q$ and $s$. The gluonic degrees of freedom are incorporated through the gluon field-strength tensor
\begin{equation}
G^{\mu\nu}=\sum_{A=1}^8 \frac{\lambda^{A}}{2} G_{A}^{\mu\nu},\
\end{equation}
where $\lambda^{A}$ are the Gell-Mann matrices generating the color SU(3) group. The interpolating current is constructed to possess the same quantum numbers as the double strange hybrid baryon under consideration and to couple to both the negative-parity ground state and the orbitally excited positive-parity state.

Having specified the interpolating current, we proceed with the construction of the hadronic representation of the correlation function. To this end, complete sets of intermediate baryonic states carrying the same quantum numbers as the current are inserted into Eq.~\eqref{eq:CorrF1}. Since the current couples to both the negative-parity ground state and the orbitally excited positive-parity state, contributions from both states must be taken into account. As a result, the correlation function on the phenomenological side can be expressed in terms of the corresponding hadronic parameters as
\begin{eqnarray}
&&\Pi ^{\mathrm{Phys}}(q) = \frac{\langle 0|\eta_{H}|\widetilde{
H }(q,\widetilde{s})\rangle \langle
\widetilde{
H }(q,\widetilde{s})|\bar{\eta}_{H}|0\rangle }{\widetilde{m}
^{2}-q^{2}}  \notag \\
&&+\frac{\langle 0|\eta_{H}|H(q,s)\rangle \langle
H (q,s)|\bar{\eta}_{H}|0\rangle }{m^{2}-q^{2}}
+\cdots ,  \label{eq:CF1/2}
\end{eqnarray}
Here, $m$ and $\widetilde{m}$ correspond to the masses of the orbitally excited positive-parity state and the negative-parity ground state of the double strange hybrid baryon, respectively, while $s$ and $\widetilde{s}$ denote their spin degrees of freedom. The ellipsis represents the contributions of higher resonances and continuum states with the same quantum numbers as the interpolating current. In deriving Eq.~\eqref{eq:CF1/2}, summation over the spin polarizations of the intermediate states is implicitly understood.

We proceed by introducing the matrix elements of the interpolating current with respect to the negative-parity ground state and the orbitally excited positive-parity state of the double strange hybrid baryon as
\begin{eqnarray}
\langle 0|\eta_{H} |\widetilde{
H }(q,\widetilde{s})\rangle  &=&\widetilde{
\lambda }\gamma _{5}u^{-}(q,\widetilde{s}),  \notag \\
\langle 0|\eta_{H} |H (q,s)\rangle  &=&\lambda u(q,s)  \label{eq:MElem}
\end{eqnarray}
where $\widetilde{\lambda}$ and $\lambda$ denote the pole residues of the negative-parity ground state and the positive-parity orbitally excited state, respectively. Using Eqs.~\eqref{eq:CF1/2} and \eqref{eq:MElem}, and performing summation over the spin polarizations of the intermediate baryons, we employ the standard relation
\begin{eqnarray}
\sum\limits_{s}u(q,s)\overline{u}(q,s) &=&\slashed q+m, 
\end{eqnarray}
to obtain
\begin{equation}
\Pi ^{\mathrm{Phys}}(q)=\frac{
\widetilde{\lambda }(\slashed q-\widetilde{m})}{\widetilde{m}^{2}-q^{2}}+\frac{\lambda (\slashed q+m)}{m^{2}-q^{2}}
+\cdots . \
\end{equation}
The Borel transformation of this expression is:
\begin{eqnarray}
\mathcal{B}\Pi ^{\mathrm{Phys}}(q) &=&\widetilde{\lambda }^{2}e^{-\frac{\widetilde{m}^{2}}{M^{2}}}(\slashed q-
\widetilde{m})  \notag \\
&+&\lambda ^{2}e^{-\frac{m^{2}}{M^{2}}}(
\slashed q+m).  \label{eq:Bor1}
\end{eqnarray}

The OPE representation of the correlation function is obtained by evaluating the same correlator in terms of quark and gluon degrees of freedom in the deep Euclidean region. Substituting the explicit form of the interpolating current into Eq.~\eqref{eq:CorrF1} and performing the Wick contractions of the quark fields, the correlation function can be expressed in terms of quark propagators and gluonic matrix elements:
\begin{eqnarray}
&&\Pi_{H}^{\mathrm{OPE}}(q)
=
-g_{s}^{2}\frac{i}{2}\epsilon_{abc}\epsilon_{a^{\prime }b^{\prime }c^{\prime }}
\int d^4x e^{iqx}\notag \\
&&\times\langle 0 |G_{\mu \nu}(x) G_{\mu^{\prime } \nu ^{\prime }}(0)|0 \rangle
(\gamma_{\nu}\gamma_{5} S^{cc^{\prime }}_{q}(x)\gamma_{5}\gamma_{\nu ^{\prime }})\notag \\
&&\times \Bigg\{
\mathrm{Tr}\Big[
\gamma_{\mu ^{\prime }}\widetilde{S}^{aa^{\prime }}_{s}(x)
\gamma_{\mu}S^{bb^{\prime }}_{s}(x)
\Big]\notag \\
&&-\mathrm{Tr}\Big[\gamma_{\mu^{\prime }}\widetilde{S}^{ba^{\prime }}_{s}(x)\gamma_{\mu}S^{ab^{\prime }}_{s}(x)\Big] \Bigg\},
\label{eq:CorrF2}
\end{eqnarray}
In Eq.~\eqref{eq:CorrF2}, $S_q(x)$ and $S_s(x)$ denote the light- and strange-quark propagators, respectively. The charge-conjugated quark propagator is defined as $\widetilde{S}(x)=CS^{T}(x)C$, and $G_{\mu\nu}$ is the gluon field-strength tensor. The explicit expression for the light-quark propagator is given by
\begin{eqnarray}
&&S_{q}^{ab}(x)=i\frac{\slashed x\delta _{ab}}{2\pi ^{2}x^{4}}-\frac{
m_{q}\delta _{ab}}{4\pi ^{2}x^{2}}-\frac{\langle \overline{q}q\rangle }{12}
\left( 1-i\frac{m_{q}}{4}\slashed x\right) \delta _{ab}  \notag \\
&&-\frac{x^{2}}{192}\langle \overline{q}g_{s}\sigma Gq\rangle \left( 1-i
\frac{m_{q}}{6}\slashed x\right) \delta _{ab}-\frac{\slashed xx^{2}g_{s}^{2}
}{7776}\langle \overline{q}q\rangle ^{2}\delta _{ab}  \notag \\
&&-\frac{ig_{s}G_{ab}^{\mu \nu }}{32\pi ^{2}x^{2}}\left[ \slashed x\sigma
_{\mu \nu }+\sigma _{\mu \nu }\slashed x\right] -\frac{x^{4}\langle
\overline{q}q\rangle \langle g_{s}^{2}G^{2}\rangle }{27648}\delta
_{ab}+\cdots.  \notag \\
&&  \label{eq:qProp}
\end{eqnarray}

Furthermore, the correlation function contains matrix elements involving two gluon field-strength tensors. To evaluate these contributions, we follow the procedure described in Ref.~\cite{Barsbay:2024vjt}. Within this framework, the gluon-field contributions are decomposed into vacuum and propagating-gluon terms. The former accounts for the interaction of gluon fields with the QCD vacuum and gives rise to condensate terms, while the latter is described by the coordinate-space gluon propagator and corresponds to the propagation of an explicit valence gluon. Both contributions are consistently incorporated within the framework of the operator product expansion (OPE).

After performing the Fourier transformation, Borel transformation, and continuum subtraction, the OPE side of the correlation function can be expressed in terms of two independent Lorentz structures as
\begin{equation}
\mathcal{B}\Pi ^{\mathrm{OPE}}(q)=\mathcal{B}\Pi
_{1}^{\mathrm{OPE}} \slashed q+\mathcal{B}\Pi
_{2}^{\mathrm{OPE}}I.
\end{equation}

The explicit expressions of the invariant amplitudes
$\mathcal{B}\Pi_{1}^{\mathrm{OPE}}$ and $\mathcal{B}\Pi_{2}^{\mathrm{OPE}}$
are given in the Appendix.

By employing the quark-hadron duality assumption, the hadronic and OPE representations of the correlation function are matched. Matching the coefficients of the Lorentz structures $\slashed{q}$ and $I$ leads to two coupled QCD sum rules, which are subsequently used to extract the masses and pole residues of the negative-parity ground state and the orbitally excited positive-parity state of the double strange hybrid baryon.
\begin{eqnarray}
\widetilde{\lambda}^{2}e^{-\frac{
\widetilde{m}^{2}}{M^{2}}}+\lambda
^{2}e^{-\frac{m^{2}}{M^{2}}}&=&\mathcal{B}\Pi _{1}^{\mathrm{OPE}},
\notag
\\
-\widetilde{\lambda}^{2}\widetilde{m}e^{-\frac{
\widetilde{m}^{2}}{M^{2}}}+\lambda
^{2}me^{-\frac{m^{2}}{M^{2}}}&=&\mathcal{B}\Pi _{2}^{\mathrm{OPE}}.
\label{eq:MFor1}
\end{eqnarray}

Solving the obtained sum rules for the hadronic parameters, the masses and pole residues of the negative-parity ground state and the orbitally excited positive-parity state are extracted. Since the correlation function contains two independent Lorentz structures, $\slashed{q}$ and $I$, the corresponding sum rules are analyzed separately for each structure. The spectroscopic parameters obtained from these two structures are then combined. The final predictions for the masses and pole residues are obtained by taking the average of the results extracted from the $\slashed{q}$ and $I$ structures.

In this procedure, the averaged quantities for the negative-parity ground state are obtained from the values extracted from the two Lorentz structures,
$\widetilde{m}_{\slashed q}$, $\widetilde{m}_{I}$ and $\widetilde{\lambda}_{\slashed q}$, $\widetilde{\lambda}_{I}$, while the corresponding quantities for the orbitally excited positive-parity state are determined from $m_{\slashed q}$, $m_{I}$ and $\lambda_{\slashed q}$, $\lambda_{I}$. This procedure is applied independently to both states, and the resulting averaged values are used as the final predictions for the masses and pole residues of the double strange hybrid baryon system.

\section{Numerical Results}
\label{sec:Numeric}

In this section, we present the numerical analysis of the QCD sum rules to determine the masses and pole residues of the double strange hybrid baryon. The input parameters of the theory, including quark masses and various vacuum condensates, are taken from the PDG and standard literature. Their numerical values are summarized in Table~\ref{tab:Param}.

\begin{table}[tbp]
\centering
\begin{tabular}{|c|c|}
\hline\hline
Parameters & Values \\ \hline\hline
$m_{s}$ & $(92.9 \pm 0.4)\,\mathrm{MeV}$ \\
$\langle \bar{q}q \rangle$ & $(-0.24 \pm 0.01)^3\,\mathrm{GeV}^3$ \\
$\langle \bar{s}s \rangle$ & $0.8\,\langle \bar{q}q \rangle$ \\
$m_{0}^{2}$ & $(0.8 \pm 0.1)\,\mathrm{GeV}^2$ \\
$\langle \bar{q} g_s \sigma G q \rangle$ & $m_{0}^{2}\langle \bar{q}q \rangle$ \\
$\langle \bar{s} g_s \sigma G s \rangle$ & $m_{0}^{2}\langle \bar{s}s \rangle$ \\
$\left\langle \frac{\alpha_s G^2}{\pi} \right\rangle$ & $(0.012 \pm 0.004)\,\mathrm{GeV}^4$ \\
\hline\hline
\end{tabular}
\caption{Input parameters used in the numerical analysis.}
\label{tab:Param}
\end{table}

After fixing the input parameters, we determine the working regions of the auxiliary parameters $M^{2}$ and $s_{0}$. These parameters are not physical observables; therefore, the physical quantities are required to exhibit stability within the chosen windows. The continuum threshold $s_{0}$ is chosen to effectively separate the lowest-lying states from higher resonances and the continuum, while the Borel parameter $M^{2}$ is constrained by the standard requirements of pole dominance and OPE convergence.

The pole contribution is defined as

\begin{equation}
\mathrm{PC}=
\frac{\Pi(M^{2},s_{0})}
{\Pi(M^{2},\infty)},
\label{eq:PC}
\end{equation}
where $\Pi(M^{2},s_{0})$ corresponds to the pole (ground-state) contribution and $\Pi(M^{2},\infty)$ represents the full phenomenological side, including both the ground-state and continuum contributions. A reliable QCD sum rule analysis requires a sufficiently large pole contribution to ensure the dominance of the lowest-lying state. In the present work, this is implemented by imposing $\mathrm{PC} \geq 0.5$, which fixes the upper bound of the Borel window.

The convergence of the operator product expansion is examined through the ratio

\begin{equation}
R(M^{2})=
\frac{\Pi^{\mathrm{DimN}}(M^{2},s_{0})}
{\Pi(M^{2},s_{0})},
\label{eq:Conv}
\end{equation}
where $\Pi^{\mathrm{DimN}}(M^{2},s_{0})$, with $\mathrm{DimN}=\mathrm{Dim}(8+9+10)$, denotes the combined contributions of the highest-dimensional nonperturbative terms included in the OPE. This ratio is used to test the convergence behavior of the OPE within the chosen Borel window. The lower bound of $M^{2}$ is fixed by requiring that the contributions of higher-dimensional condensates remain sufficiently suppressed, ensuring the stability and convergence of the OPE. It should be noted that, due to the presence of explicit gluonic degrees of freedom in the interpolating current for the double strange hybrid baryon, particular care is required in the stability analysis of the sum rules.

The analysis is performed using two independent Lorentz structures, $\slashed{q}$ and $I$. The results obtained from these structures are found to be consistent within uncertainties. To reduce possible systematic effects associated with the choice of Lorentz structure, we take the average of the results for both the ground state and the first excited state.

The working regions of the auxiliary parameters used in the analysis are summarized in Table~\ref{tab:results}.

\begin{widetext}

\begin{table}[tbp]
\centering
\begin{tabular}{|c|c|c|c|c|}
\hline\hline
State & $M^2~(\mathrm{GeV}^2)$ & $s_0~(\mathrm{GeV}^2)$  & Mass $(\mathrm{MeV})$ & Pole residue $(\mathrm{GeV}^5)$\\
\hline
Ground state ($1/2^{-}$) & $[2-4]$ & $[3.5-4.5]$ & $(1593.44\pm 130.29)$ & $(2.57\pm 0.40) 
\times10^{-3}$ \\
Excited state ($1/2^{+}$) & $[2-4]$ & $[4.5-5.5]$ & $(1897.47\pm 124.44)$ & $(2.88\pm 0.62) 
\times10^{-3}$ \\
\hline\hline
\end{tabular}
\caption{Working regions of the auxiliary parameters and averaged values of the masses and pole residues obtained from the $\slashed{q}$ and $I$ Lorentz structures for the double strange hybrid baryon states.}
\label{tab:results}
\end{table}

\end{widetext}

For clarity, the dependence of the masses and pole residues on the Borel parameter $M^{2}$ and the continuum threshold $s_{0}$ is analyzed separately. First, the ground-state mass is studied as a function of $M^{2}$ and $s_{0}$, followed by its pole residue. The same procedure is then applied to the first orbitally excited state.

The dependence of the ground-state mass on $M^{2}$ and $s_{0}$ is illustrated in Fig.~\ref{fig:gr_mass}, while the corresponding pole residue is shown in Fig.~\ref{fig:gr_res}. Similarly, Figs.~\ref{fig:exc_mass} and \ref{fig:exc_res} present the results for the first excited state.

\begin{widetext}

\begin{figure}[h!]
\begin{center}
\includegraphics[width=8cm,height=6cm]{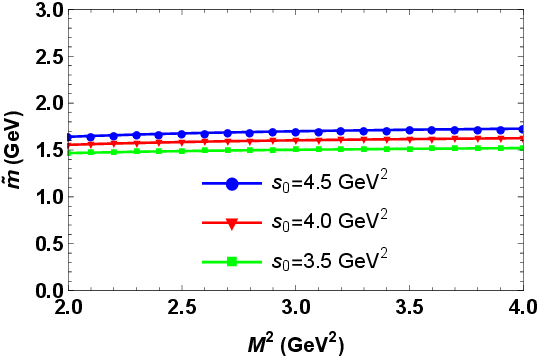}\,\,
\includegraphics[width=8cm,height=6cm]{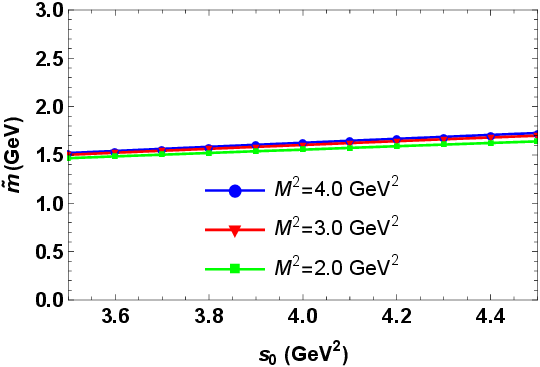}
\end{center}
\caption{Dependence of the ground-state mass on the Borel parameter $M^{2}$ (left panel) and the continuum threshold $s_{0}$ (right panel), obtained from the averaged results of the $\slashed{q}$ and $I$ Lorentz structures.}
\label{fig:gr_mass}
\end{figure}

\begin{figure}[h!]
\begin{center}
\includegraphics[width=8cm,height=6cm]{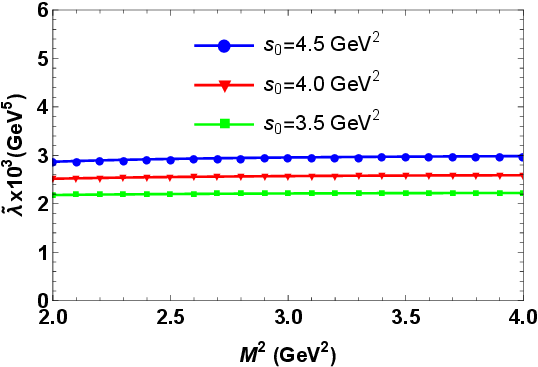}\,\,
\includegraphics[width=8cm,height=6cm]{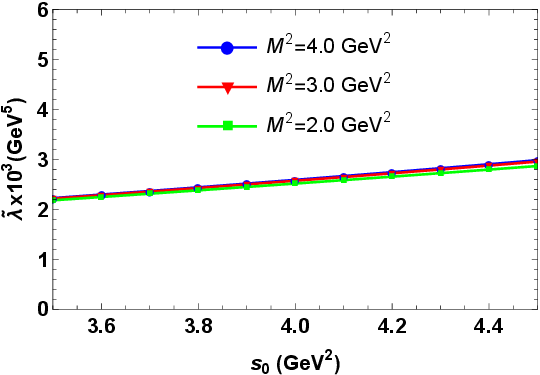}
\end{center}
\caption{Dependence of the ground-state pole residue on the Borel parameter $M^{2}$ (left panel) and the continuum threshold $s_{0}$ (right panel), obtained from the averaged $\slashed{q}$ and $I$ Lorentz structures.}
\label{fig:gr_res}
\end{figure}

\begin{figure}[h!]
\begin{center}
\includegraphics[width=8cm,height=6cm]{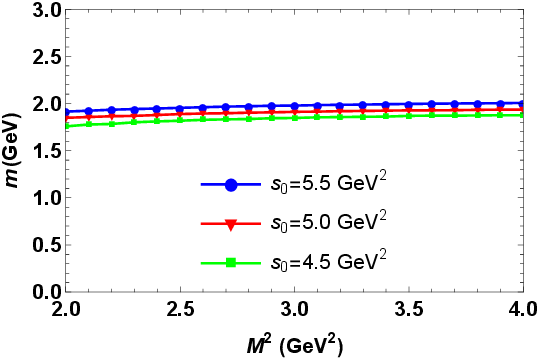}\,\,
\includegraphics[width=8cm,height=6cm]{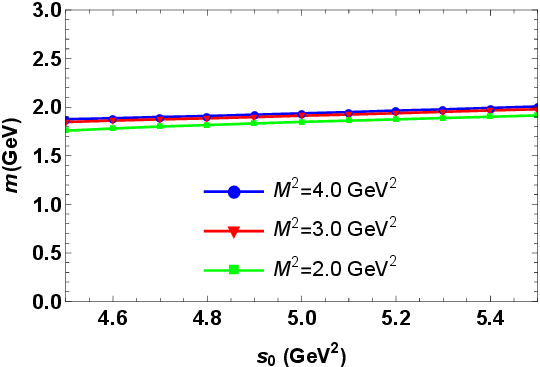}
\end{center}
\caption{Dependence of the first orbitally excited state mass on the Borel parameter $M^{2}$ (left panel) and the continuum threshold $s_{0}$ (right panel), obtained from the averaged $\slashed{q}$ and $I$ Lorentz structures.}
\label{fig:exc_mass}
\end{figure}

\begin{figure}[h!]
\begin{center}
\includegraphics[width=8cm,height=6cm]{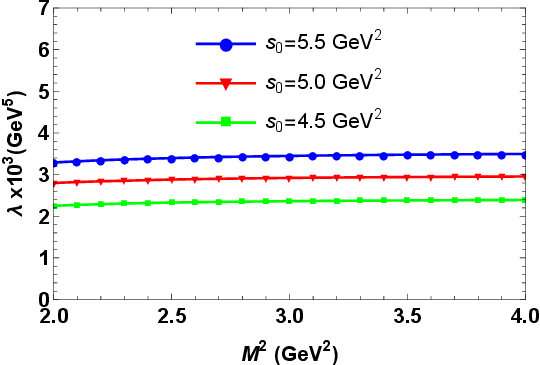}\,\,
\includegraphics[width=8cm,height=6cm]{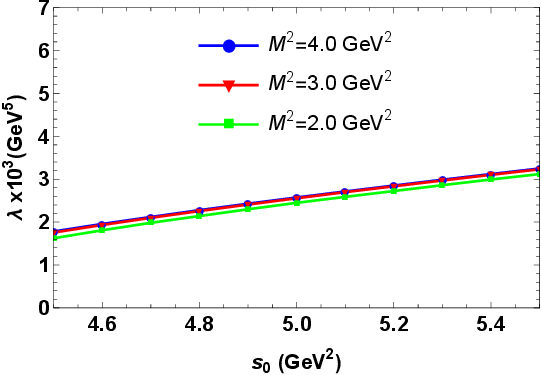}
\end{center}
\caption{Dependence of the first orbitally excited state pole residue on the Borel parameter $M^{2}$ (left panel) and the continuum threshold $s_{0}$ (right panel), obtained from the averaged $\slashed{q}$ and $I$ Lorentz structures.}
\label{fig:exc_res}
\end{figure}

\end{widetext}

The analysis shows that the results are stable under variations of the Borel parameter $M^{2}$ and the continuum threshold $s_{0}$ within the working regions. Moreover, the results obtained from the two Lorentz structures are consistent within theoretical uncertainties.

Finally, the averaged results obtained from the $\slashed{q}$ and $I$ Lorentz structures are adopted as the final predictions for the masses and pole residues of both the negative-parity ground state and the first orbitally excited positive-parity state of the double strange hybrid baryon.

\section{Conclusions}

\label{sec:Dis} 

In this work, we have investigated the spectroscopic properties of the double strange hybrid baryon with quark content $ssqg$ within the framework of QCD sum rules. The two-point correlation function has been constructed using an interpolating current with explicit gluonic degrees of freedom and analyzed by considering two independent Lorentz structures, $\slashed{q}$ and $I$. The operator product expansion has been performed by including nonperturbative contributions up to dimension ten, and the corresponding sum rules have been derived for both structures.

The working regions of the Borel parameter and continuum threshold have been determined by requiring the convergence of the OPE and the dominance of the pole contribution. The numerical analysis shows that the extracted quantities exhibit stable behavior within the selected working windows. The results obtained from the two Lorentz structures are consistent within theoretical uncertainties, and the final predictions have been obtained by averaging the corresponding results.

Our results show that the negative-parity ground state of the double strange hybrid baryon has a mass around $1.6~\mathrm{GeV}$, while the first orbitally excited positive-parity state appears around $1.9~\mathrm{GeV}$. The corresponding pole residues have also been determined, providing information about the coupling of the interpolating current to these states.

The obtained mass predictions can be compared with previous QCD sum rule studies of hybrid baryons. In Ref.~\cite{Wang:2024lnv}, the spectroscopic properties of light hybrid baryons were investigated using QCD sum rules with local $qqqg$ interpolating currents, and the lowest-lying hybrid baryon states were predicted to have masses around $2~\mathrm{GeV}$. Similarly, Ref.~\cite{Kisslinger:1995yw} presented an early QCD sum rule analysis of light hybrid baryons and estimated the mass of the lowest hybrid state to be approximately $1.5~\mathrm{GeV}$. The present predictions for the double strange $ssqg$ hybrid baryon lie within a similar energy scale. The differences with respect to previous results can be attributed to the different flavor structure, the inclusion of strange quark degrees of freedom, and the specific interpolating current employed in the present analysis.

The present study provides theoretical predictions for the low-lying spectrum of the double strange hybrid baryon and contributes to a better understanding of hybrid baryon structures within the QCD sum rule framework. These results may serve as useful references for future theoretical studies and experimental searches for exotic baryonic states.

\appendix*

\begin{widetext}

\section{ The OPE side of the correlation function}

\renewcommand{\theequation}{\Alph{section}.\arabic{equation}} \label{sec:App}

In this appendix, we provide the explicit forms of the functions $\mathcal{B}\Pi^{\mathrm{OPE}}_1(q)$ and $\mathcal{B}\Pi^{\mathrm{OPE}}_2(q)$ that are employed in the mass sum rules:

\begin{eqnarray}
\mathcal{B}\Pi_{1}^{\mathrm{OPE}}(q)
&=&
\int_{m_{s}^{2}}^{s_0} ds\, e^{-s/M^2}\Theta(N)\,\rho_1(s)
\nonumber\\
&&
+\frac{g_s^2 m_0^4 m_{s}^2 \langle \bar{s}s\rangle^2}{1152M^2\pi^2}
+\frac{\langle \alpha _{s}G^{2}/\pi
\rangle g_s^2 m_{s}^2 \langle \bar{q}q\rangle^2}{2592M^2},
\label{eq:Pi1QCD}
\end{eqnarray}

\begin{eqnarray}
\rho_1(s)
&=&
-\frac{g_s^2 s^3(12m_{s}^2+s)}{245760\pi^6}-\frac{\langle \alpha _{s}G^{2}/\pi
\rangle g_s^2 s^2}{24576\pi^4}-\frac{\langle \alpha _{s}G^{2}/\pi
\rangle m_{s}^2 s}{256\pi^2} \nonumber\\
&&
-\frac{g_s^2 m_0^2 m_{s} s \langle \bar{s}s\rangle}{768\pi^4}
-\frac{g_s^2
\Big[
54\pi^2(3m_{s}^2+2s)\langle \bar{s}s\rangle^2
+g_s^2(
3m_{s}^2\langle \bar{q}q\rangle^2
+2s\langle \bar{q}q\rangle^2
+4s\langle \bar{s}s\rangle^2)
\Big]}{10368\pi^4}
\nonumber\\
&&
+\frac{\langle \alpha _{s}G^{2}/\pi
\rangle g_s^2 m_{s} \langle \bar{s}s\rangle}{1536\pi^2}
+\frac{\langle \alpha _{s}G^{2}/\pi
\rangle m_{s}\langle \bar{s}s\rangle}{32}
+\frac{g_s^2 m_0^2 \langle \bar{s}s\rangle^2}{64\pi^2}
\end{eqnarray}
and
\begin{eqnarray}
\mathcal{B}\Pi_{2}^{\mathrm{OPE}}(q)
&=&
\int_{m_{s}^{2}}^{s_0} ds\, e^{-s/M^2} \Theta(N)\,\rho_2(s)
\nonumber\\
&&
+\frac{1}{64}\langle \alpha _{s}G^{2}/\pi
\rangle
m_0^2 m_{s}^2 \langle \bar{q}q\rangle
-\frac{1}{576}\langle \alpha _{s}G^{2}/\pi
\rangle
g_s^2 m_{s} \langle \bar{q}q\rangle \langle \bar{s}s\rangle
\nonumber\\
&&
-\frac{g_s^2 m_0^4 m_{s} \langle \bar{q}q\rangle \langle \bar{s}s\rangle}{384\pi^2}
+\frac{1}{8}\langle \alpha _{s}G^{2}/\pi
\rangle m_{s}\pi^2 \langle \bar{q}q\rangle \langle \bar{s}s\rangle,
\label{eq:Pi2QCD}
\end{eqnarray}

\begin{eqnarray}
\rho_2(s)
&=&
\frac{g_s^2 \langle \bar{q}q\rangle s^3}{3072\pi^4}
-\frac{g_s^2 m_0^2 s^2 \langle \bar{q}q\rangle}{1024\pi^4}
+\frac{g_s^2 m_{s} s \langle \bar{q}q\rangle \langle \bar{s}s\rangle}{32\pi^2}
\nonumber\\
&&
+\frac{\langle \alpha _{s}G^{2}/\pi
\rangle g_s^2 s \langle \bar{q}q\rangle}{768\pi^2}
+\frac{1}{64}\langle \alpha _{s}G^{2}/\pi
\rangle \langle \bar{q}q\rangle (4m_{s}^2+s)
\nonumber\\
&&
-\frac{5 g_s^2 m_0^2 m_{s} \langle \bar{q}q\rangle \langle \bar{s}s\rangle}{192\pi^2}
-\frac{\langle \bar{q}q\rangle
\left(27 g_s^2 m_0^2 \langle \alpha _{s}G^{2}/\pi
\rangle -128 g_s^4 \langle \bar{s}s\rangle^2\right)}{82944\pi^2}
\nonumber\\
&&
-\frac{1}{128}\langle \alpha _{s}G^{2}/\pi
\rangle m_0^2 \langle \bar{q}q\rangle.
\end{eqnarray}

Here, $\Theta ({N)}$ denotes the unit step function with argument $N=s$. In the present analysis, we work in the limit $m_u=m_d=0$, while retaining a nonzero strange quark mass $m_s\neq0$.

\end{widetext}

\renewcommand{\theequation}{\Alph{section}.\arabic{equation}} \label{sec:App}


\end{document}